\documentclass[showpacs,aps,graphicx,twocolumn]{revtex4}
\usepackage{graphicx}

\begin{document}
\title{Multiparty quantum state sharing of an arbitrary
two-particle state with Einstein-Podolsky-Rosen pairs}

\author{ Fu-Guo Deng,$^{1,2,3}$ Xi-Han Li,$^{1,2}$ Chun-Yan Li,$^{1,2}$
Ping Zhou,$^{1,2}$ and Hong-Yu Zhou$^{1,2,3}$ }
\address{$^1$ The Key Laboratory of Beam Technology and Material
Modification of Ministry of Education, Beijing Normal University,
Beijing 100875,
People's Republic of China\\
$^2$ Institute of Low Energy Nuclear Physics, and Department of
Material Science and Engineering, Beijing Normal University,
Beijing 100875,
People's Republic of China\\
$^3$ Beijing Radiation Center, Beijing 100875,  People's Republic
of China}
\date{\today }

\begin{abstract}
A scheme for multiparty quantum state sharing of an arbitrary
two-particle state is presented with Einstein-Podolsky-Rosen
pairs. Any one of the $N$ agents has the access to regenerate the
original state with two local unitary operations if he
collaborates with the other agents, say the controllers. Moreover,
each of the controllers is required to take only a product
measurement $\sigma_x \otimes \sigma_x$ on his two particles,
which makes this scheme more convenient for the agents in the
applications on a network than others. As all the quantum source
can be used to carry the useful information, the intrinsic
efficiency of qubits approaches the maximal value. With a new
notation for the multipartite entanglement, the sender need only
publish two bits of classical information for each measurement,
which reduces the information exchanged largely.

\end{abstract}
\pacs{03.67.Hk, 03.67.Dd, 03.65.Ud, 89.70.+c} \maketitle

Quantum secret sharing (QSS), an important branch of quantum
communication, is the generalization of classical secret sharing
\cite{Blakley} into quantum scenario and has attracted a lot of
attention
\cite{HBB99,KKI,QSSS,longqss,cleve,nascimento,Peng,Lance,GuoQSSPLA,dengPLA,Hsu,TZG,AMLance}.
There are three main goals in QSS: (1) it is used to distribute a
private key among many users
\cite{HBB99,KKI,QSSS,longqss,GuoQSSPLA,dengPLA,Hsu}, similar to
quantum key distribution (QKD) \cite{QKDs}; (2) it is a tool for
sharing a classical secret directly
\cite{HBB99,KKI,longqss,cleve,nascimento}, similar to quantum
secure direct communication (QSDC) \cite{QSDC,two-step}; (3) it
provides a secure way for sharing a quantum information (an
unknown quantum state) \cite{cleve,nascimento,Peng,Lance}, similar
to the controlled teleportation
\cite{ControlledTele,ControlledTele2,ControlledTele3}. Most
existing QSS schemes are focused on creating a private key among
several parties or splitting a classical secret. For example, an
original QSS scheme \cite{HBB99} was proposed by Hillery,
Bu\v{z}ek, and Berthiaume (HBB) in 1999 by using a three-particle
or a four-particle Greenberger-Horne-Zeilinger (GHZ) state for
distributing a private key among some agents and sharing a
classical information.

In recent, an interest work was done by Li et al. \cite{Peng} for
sharing an unknown single qubit with a multipartite joint
measurement. In their QSS protocol, the sender splits a qubit into
$m$ pieces for the $m$ agents with $m$ Einstein-Podolsky-Rosen
(EPR) pairs, and any one of the agents can obtain the qubit with
the help of the other agents. In 2004, Lance et al. \cite{Lance}
named the branch of quantum secret sharing for quantum information
"quantum-state sharing" (QSTS). By far, there are no models for
sharing an arbitrary multipartite state. In this paper, we will
present a way for sharing an arbitrary two-particle state with
$2N$ EPR pairs. Any one in the $N$ agents can regenerate the
original state when he collaborates with the others, say the
controllers. Moreover, the controllers need only perform the
single-particle measurements on their particles, and the receiver
can reconstruct the original state with two local unitary
operations.

The basic idea of this QSTS for an arbitrary  two-particle state
with two agents is shown in Fig.1. Alice is the sender, Bob and
Charlie are the two agents. Suppose that the unknown arbitrary
two-particle state is described as
\begin{eqnarray}
\vert \Phi\rangle_{xy}=\alpha\vert 00\rangle_{xy} + \beta \vert
01\rangle_{xy} + \gamma\vert 10\rangle_{xy} + \delta\vert
11\rangle_{xy},\label{unknownstate}
\end{eqnarray}
where $x$ and $y$ are the two particles in the state $\vert
\Phi\rangle_{xy}$, and
\begin{eqnarray}
\vert \alpha\vert^2 + \vert \beta\vert^2 + \vert \gamma\vert^2 +
\vert \delta\vert^2 =1.
\end{eqnarray}
At first, Alice shares the four EPR pairs $a_1b_1$, $c_1d_1$,
$a_2b_2$ and $c_2d_2$ with Bob and Charlie, respectively. Here
$a_1$ and $b_1$ are the two particles in an EPR pair, and similar
notations for the other EPR pairs. An EPR pair is in one of the
four Bell states shown as follows \cite{book}:
\begin{eqnarray}
\left\vert \psi ^{\pm}\right\rangle =\frac{1}{\sqrt{2}}(\left\vert
0\right\rangle\left\vert 1\right\rangle\pm\left\vert
1\right\rangle \left\vert 0\right\rangle), \label{EPR12}\nonumber\\
\left\vert \phi ^{\pm}\right\rangle =\frac{1}{\sqrt{2}}(\left\vert
0\right\rangle\left\vert 0\right\rangle\pm\left\vert
1\right\rangle\left\vert 1\right\rangle), \label{EPR34}
\end{eqnarray}
where $\vert 0\rangle$ and $\vert 1\rangle$  are the eigenvectors
of the operator $\sigma_z$. Without loss of generalization, we
assume that all the EPR pairs are originally in the entangled
state $\vert \phi^+\rangle=\frac{1}{\sqrt{2}}(\left\vert
0\right\rangle\left\vert 0\right\rangle +\left\vert
1\right\rangle\left\vert 1\right\rangle)$.

\begin{figure}[!h]
\begin{center}
\includegraphics[width=8cm,angle=0]{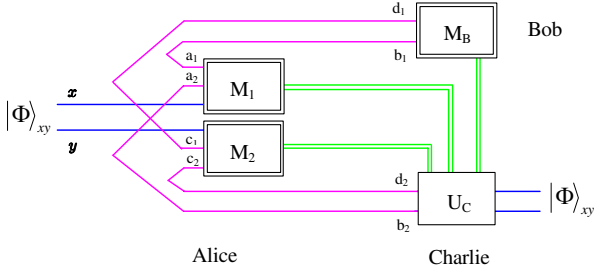} \label{fig1}
\caption{(Color online) Multiparty quantum secret sharing for an
arbitrary two-particle state with two agents. The single lines
denote qubits, double lines denote classical data, similar to Ref.
\cite{Leung}. $M_1$, $M_2$ are the GHZ-state joint measurements on
the particles $xa_1a_2$ and $yc_1c_2$, respectively; $M_B$ is the
product measurement $\sigma_x\otimes \sigma_x$ on the particles
$b_1d_1$. }
\end{center}
\end{figure}

Before the measurement, the state of the composite quantum system
composed of the ten particles is
\begin{eqnarray}
\vert \Psi\rangle_{s}&\equiv& \vert \Phi\rangle_{xy} \vert
\phi^+\rangle_{a_1b_1}  \vert \phi^+\rangle_{c_1d_1}  \vert
\phi^+\rangle_{a_2b_2}  \vert \phi^+\rangle_{c_2d_2}.
\end{eqnarray}
Alice performs the three-particle GHZ state joint measurement
$M_1$ on the particles $x$, $a_1$ and $a_2$ first, and then the
$M_2$ on the particles $y$, $c_1$ and $c_2$. Bob takes the product
measurement $M_B=\sigma_x\otimes \sigma_x$ on the particles
$b_1d_1$, and then Charlie can recover the original state
$\vert\Phi\rangle_{xy}$ with two local unitary operations
$U_C=U_b\otimes U_d$ according to the results obtained by Alice
and Bob, see Fig.1.

Let us use an example to demonstrate the principle of this QSTS
protocol with one controller. First, we introduce a new notation
for the three-particle GHZ states.
\begin{eqnarray}
\vert G_{ij+}\rangle =\frac{1}{\sqrt{2}} (\vert 0ij\rangle + \vert
1\bar{i}\bar{j}\rangle),\,    \vert G_{ij-}\rangle
=\frac{1}{\sqrt{2}} (\vert 0ij\rangle - \vert
1\bar{i}\bar{j}\rangle),
\end{eqnarray}
where $i,j\in \{0, 1\}$, $\bar{i}=1-i$ and $\bar{j}=1-j$.

Suppose Alice gets the results $R_{xa_1a_2}=R_{yc_1c_2}=\vert
G_{00+} \rangle$, which will occurs with the probability
$\frac{1}{8}\times\frac{1}{8}=\frac{1}{64}$, then the state of the
subsystem with the particles $b_1$, $d_1$, $b_2$ and $d_2$ becomes
\begin{eqnarray}
\vert \Psi\rangle_{sub} =\alpha\vert 00\rangle_{b_1d_1} \vert
00\rangle_{b_2d_2} + \beta\vert 01\rangle_{b_1d_1} \vert
01\rangle_{b_2d_2}\nonumber\\
 + \gamma\vert 10\rangle_{b_1d_1}
\vert 10\rangle_{b_2d_2} + \delta \vert 11\rangle_{b_1d_1} \vert
11\rangle_{b_2d_2}.
\end{eqnarray}
That is, the information of the state $\vert \Phi\rangle_{xy}$ is
transferred to the state of the subsystem shared between Bob and
Charlie. If they want to recover the quantum information $\vert
\Phi\rangle_{xy}$, one of them performs $\sigma_x \otimes
\sigma_x$ on his/her two particles and the other takes two local
unitary operations on the two particles remained according to the
information provided by the first one. For example, let us assume
that Bob performs the $\sigma_x \otimes \sigma_x$ measurement on
his two particles, and Charlie will reconstruct the original state
when she collaborates with Bob. We can rewrite the state $\vert
\Psi\rangle_{sub}$ as
\begin{eqnarray}
&&\vert \Psi\rangle_{sub} =\frac{1}{2}[\vert +x\rangle_{b_1}\vert
+x\rangle_{d_1}(\alpha\vert 00\rangle_{b_2d_2} + \beta\vert
01\rangle_{b_2d_2}\nonumber\\
&&+ \gamma\vert 10\rangle_{b_2d_2} + \delta\vert
11\rangle_{b_2d_2})+ \vert +x\rangle_{b_1}\vert
-x\rangle_{d_1}(\alpha\vert 00\rangle_{b_2d_2} \nonumber\\
&&- \beta\vert 01\rangle_{b_2d_2} + \gamma\vert 10\rangle_{b_2d_2}
- \delta\vert 11\rangle_{b_2d_2}) +\vert -x\rangle_{b_1}\vert
+x\rangle_{d_1}\nonumber\\
&&(\alpha\vert 00\rangle_{b_2d_2} + \beta\vert 01\rangle_{b_2d_2}
- \gamma\vert
10\rangle_{b_2d_2} - \delta\vert 11\rangle_{b_2d_2})\nonumber\\
&&+\vert -x\rangle_{b_1}\vert -x\rangle_{d_1}(\alpha\vert
00\rangle_{b_2d_2} - \beta\vert 01\rangle_{b_2d_2} - \gamma\vert
10\rangle_{b_2d_2} \nonumber\\
&&+ \delta\vert 11\rangle_{b_2d_2})],
\end{eqnarray}
where $\vert +x\rangle=\frac{1}{\sqrt{2}}(\vert 0\rangle + \vert
1\rangle)$ and $\vert -x\rangle=\frac{1}{\sqrt{2}}(\vert 0\rangle
- \vert 1\rangle)$ are the two eigenstates of the measuring basis
$\sigma_x$. Provided that Bob agrees to cooperate with Charlie,
Charlie can recover the unknown state by performing the unitary
operations $U_0\otimes U_0 $, $U_0\otimes U_1 $, $U_1\otimes U_0
$, and $U_1\otimes U_1 $ on the particles $b_2$ and $d_2$ if the
outcomes obtained by Bob are $\vert +x\rangle_{b_1}\vert
+x\rangle_{b_2}$, $\vert +x\rangle_{b_1}\vert -x\rangle_{b_2}$,
$\vert -x\rangle_{b_1}\vert +x\rangle_{b_2}$ and $\vert
-x\rangle_{b_1}\vert -x\rangle_{b_2}$, respectively. Here
$U_0\equiv I$, $U_1\equiv \sigma_z$, $U_2\equiv \sigma_x$ and
$U_3\equiv i\sigma_y$, and I is the identity matrix and $\sigma_i$
$(i=x,y,z)$ are the Pauli matrices.

For the other cases, the relation between the results of the
measurements done by Alice and Bob and the local unitary
operations with which Charlie reconstructs the unknown quantum
information $\vert \Phi\rangle_{xy}$ is shown in Table I. Here
$V_{xa_1a_2}$ and $V_{yc_1c_2}$ represents the bit value of the
results of the GHZ state joint measurements on $xa_1a_2$ and
$yc_1c_2$, respectively. Define
\begin{eqnarray}
V_{\vert G_{ij\pm}\rangle}\equiv j,\,\,\,\,\,\,\,\, P_{\vert
G_{ij\pm}\rangle}\equiv \pm, \,\,\,\,\,\,\, P_{\vert \pm
x\rangle}\equiv \pm
\end{eqnarray}
where $i,j\in \{0,1\}$. In detail, $V_{xa_1a_2}=1$ and
$P_{xa_1a_2}= - $ if the result of the three-particle joint
measurement on particles $xa_1a_2$ is $R_{xa_1a_2}=\vert
G_{01-}\rangle$ or $R_{xa_1a_2}=\vert G_{11-}\rangle$; $P_{b_1}=-$
when $R_{b_1}=\vert -x\rangle$. $U_i\otimes U_j$ means that
Charlie performs $U_i$ and $U_j$ on the two particles $b_2$ and
$d_2$, respectively, here $i,j=0,1,2,3$.

Table I shows that the unknown state $\vert \Phi\rangle_{xy}$ can
be shared by Bob and Charlie completely, and they can reconstruct
the state with two single-qubit measurements along the
$x-$direction and two local unitary operations. They need not do
Bell state measurement on the particles, which makes this QSTS
protocol more convenient for the agents than that in Ref.
\cite{ControlledTele3}. Moreover, Alice needs only to publish two
bits of classical information for her agents to recover the state
$\vert \Phi\rangle_{xy}$.

It is straightforwardly to generalize this QSTS scheme to the case
with $N$ agents, say Bob$_i$ ($i=1,2,...,N-1$) and Charlie. As the
symmetry, we still assume that Charlie is the agent who will
recover the unknown state with the help of the $N-1$ controllers,
Bob$_i$. For the end, Alice should share $2N$ EPR pairs $\vert
\psi \rangle_{a_ib_i}$ and $\vert \psi \rangle_{c_id_i}$,
($i=1,2,\ldots,N$) with the $N$ agents. In this time, the state of
the composite quantum system is
\begin{eqnarray}
\vert \Phi\rangle_{S} \equiv \vert \Phi\rangle_{xy}
\prod_{i=1}^{N}\otimes \vert \phi^+\rangle_{a_ib_i} \otimes \vert
\phi^+\rangle_{c_id_i}.
\end{eqnarray}
Define a set of orthogonal vectors as
\begin{eqnarray}
\vert G_{\underbrace{ij\ldots k}_N+}\rangle =\frac{1}{\sqrt{2}}
(\vert 0 \underbrace{ij\ldots k}_N\rangle + \vert 1
\underbrace{\bar{i}\bar{j}\ldots \bar{k}}_N\rangle)\nonumber\\
\vert G_{\underbrace{ij\ldots k}_N-}\rangle =\frac{1}{\sqrt{2}}
(\vert 0 \underbrace{ij\ldots k}_N\rangle - \vert 1
\underbrace{\bar{i}\bar{j}\ldots \bar{k}}_N\rangle),
\end{eqnarray}
where $i,j,k\in \{0, 1\}$, $\bar{i}, \bar{j}$ and $\bar{k}$ are
the counterparts of the binary numbers $i,j$ and $k$,
respectively.

In the quantum communication, Alice performs first the joint
measurement on the $N+1$ particles $x$, $a_1$, $\ldots$, and
$a_N$, then on the $N+1$ particles $y$, $c_1$, $\ldots$, and
$c_N$. When the agents want to reconstruct the unknown state
$\vert \Phi \rangle_{xy}$, each of the controllers, Bob$_i$
performs $\sigma_x \otimes \sigma_x$ on his two particles $b_i$
and $d_i$, i.e.,
\begin{eqnarray}
\vert \Phi\rangle_{S} =\Psi_{xa_1\ldots a_{N}}\otimes
\Psi_{yc_1\ldots c_{N}}\otimes(\prod_{i=1}^{N} \Psi_{b_i})\otimes
(\prod_{i=1}^{N} \Psi_{d_i}),
\end{eqnarray}
where $\Psi_{xa_1\ldots a_{N}},  \Psi_{yc_1\ldots c_{N}} \in$ $
\{\vert G_{\underbrace{ij\ldots k}_N+}\rangle,$ $\vert
G_{\underbrace{ij\ldots k}_N-}\rangle\}$ are the results of the
joint measurements done by Alice.  In more detail, the state of
the quantum system (without being normalized) can be rewritten as
\begin{eqnarray}
\vert \Psi\rangle_{S} && = \sum_{i,j,\ldots,k \atop
m,n,\ldots,l}\{\vert G_{ij \ldots k+}\rangle \vert G_{mn \ldots
l+}\rangle (\alpha\vert ij \ldots k\rangle \vert mn \ldots
l\rangle\nonumber\\
&& + \beta\vert ij \ldots k\rangle \vert
\bar{m} \bar{n} \ldots \bar{l}\rangle + \gamma\vert \bar{i}
\bar{j} \ldots \bar{k}\rangle \vert mn \ldots l\rangle \nonumber\\
&& + \delta\vert \bar{i} \bar{j} \ldots \bar{k}\rangle \vert
\bar{m} \bar{n} \ldots \bar{l}\rangle) + \vert G_{ij \ldots
k+}\rangle \vert G_{mn \ldots l-}\rangle \nonumber\\
&& (\alpha\vert ij \ldots k\rangle \vert mn \ldots l\rangle -
\beta\vert ij \ldots k\rangle \vert \bar{m} \bar{n} \ldots
\bar{l}\rangle \nonumber\\
&& + \gamma\vert \bar{i} \bar{j} \ldots \bar{k}\rangle \vert mn
\ldots l\rangle - \delta\vert \bar{i} \bar{j} \ldots
\bar{k}\rangle \vert \bar{m} \bar{n} \ldots \bar{l}\rangle) \nonumber\\
&& + \vert G_{ij \ldots k-}\rangle \vert G_{mn \ldots l+}\rangle
(\alpha\vert ij \ldots k\rangle \vert mn \ldots l\rangle \nonumber\\
&& + \beta\vert ij \ldots k\rangle \vert \bar{m} \bar{n} \ldots
\bar{l}\rangle - \gamma\vert \bar{i} \bar{j} \ldots \bar{k}\rangle
\vert mn \ldots l\rangle\nonumber\\
&& - \delta\vert \bar{i} \bar{j} \ldots \bar{k}\rangle \vert
\bar{m} \bar{n} \ldots \bar{l}\rangle) + \vert G_{ij \ldots
k-}\rangle \vert G_{mn \ldots l-}\rangle
\nonumber\\
&& (\alpha\vert ij \ldots k\rangle \vert mn \ldots l\rangle -
\beta\vert ij \ldots k\rangle \vert \bar{m} \bar{n} \ldots
\bar{l}\rangle\nonumber\\
&& - \gamma\vert \bar{i} \bar{j} \ldots \bar{k}\rangle \vert mn
\ldots l\rangle + \delta\vert \bar{i} \bar{j} \ldots
\bar{k}\rangle \vert \bar{m} \bar{n} \ldots \bar{l}\rangle)\},
\end{eqnarray}
where $\vert i\rangle=\frac{1}{\sqrt{2}}[\vert +x
\rangle+(-1)^i\vert -x\rangle]$, $\{i,j,\ldots,k,m,n,\ldots,l \}$
are $2N$ binary numbers, and $\bar{m}$ is the counterpart of $m$,
i.e., $\bar{m}=1-m$. As the symmetry, the measurements done by the
controllers can be expressed by the operation $M$,
\begin{eqnarray}
M = [(\langle +x\vert)^{N-1-t}(\langle
-x\vert)^{t}]_1\otimes[(\langle +x\vert)^{N-1-q}(\langle
-x\vert)^{q}]_2,\nonumber\\
\end{eqnarray}
where $[(\langle +x\vert)^{N-1-t}(\langle -x\vert)^{t}]_1$ is the
measurement operation related to the state of the quantum
subsystem $b_i$ (i.e, $\prod_{i=1}^{N} \Psi_{b_i}$), and
$[(\langle +x\vert)^{N-1-q}(\langle -x\vert)^{q}]_2$ is related to
$d_i$, $t$  and $q$ are the numbers that the controllers obtain
the result $\vert -x\rangle$ when they measure the particle $b_i$
and $d_i$, respectively. After the measurements done by Alice and
the $N-1$ controllers, the relation between the state of the
particles $b_Nd_N$ and the results of the measurements can be
expressed as:
\begin{eqnarray}
M\vert \Psi\rangle_{S} &=&\sum_{i,j,\ldots,k \atop
m,n,\ldots,l}\{\vert G_{ij \ldots k+}\rangle \vert G_{mn \ldots
l+}\rangle \otimes e^{-\theta_1}(\alpha\vert  kl\rangle\nonumber\\
&+& (-1)^{q} \beta\vert k\bar{l}\rangle + (-1)^{t}\gamma\vert
\bar{k} l\rangle
+ (-1)^{t+q}\delta\vert \bar{k} \bar{l}\rangle)\nonumber\\
&+&\vert G_{ij \ldots k+}\rangle \vert G_{mn \ldots l-}\rangle
\otimes e^{-\theta_2}(\alpha\vert  kl\rangle \nonumber\\
&+& (-1)^{q+1} \beta\vert k\bar{l}\rangle + (-1)^{t}\gamma\vert
\bar{k} l\rangle
+ (-1)^{t+q+1}\delta\vert \bar{k} \bar{l}\rangle)\nonumber\\
&+&\vert G_{ij \ldots k-}\rangle \vert G_{mn \ldots l+}\rangle
\otimes e^{-\theta_3}(\alpha\vert  kl\rangle \nonumber\\
&+& (-1)^{q} \beta\vert k\bar{l}\rangle + (-1)^{t+1}\gamma\vert
\bar{k} l\rangle
+ (-1)^{t+q+1}\delta\vert \bar{k} \bar{l}\rangle)\nonumber\\
&+&\vert G_{ij \ldots k-}\rangle \vert G_{mn \ldots l-}\rangle
\otimes e^{-\theta_4}(\alpha\vert  kl\rangle \nonumber\\
&+& (-1)^{q+1} \beta\vert k\bar{l}\rangle + (-1)^{t+1}\gamma\vert
\bar{k} l\rangle + (-1)^{t+q}\delta\vert \bar{k}
\bar{l}\rangle)\},\nonumber\\
\end{eqnarray}
where $e^{-\theta_i}$ $(i=1,2,3,4)$ is an integer phase related to
the state of quantum system $b_Nd_N$, $\Psi_{b_Nd_N}$, and it does
not affect the result of the final state $\Psi_{b_Nd_N}$ after all
the measurements are completed.

Similar to the notations discussed above, we define
\begin{eqnarray}
V_{\vert G_{ij\ldots k\pm}\rangle}\equiv k,\,\,\,\,\,\,\,\,
P_{\vert G_{ij\ldots k\pm}\rangle}\equiv \pm.
\end{eqnarray}
The relation between the results of the measurements and the local
unitary operations with which Charlie reconstructs the unknown
quantum information is as same as that in Table I with just a
little modification. That is, $V_{xa_1a_2}$, $V_{yc_1c_2}$,
$P_{xa_1a_2}\otimes P_{b_1}$ and $P_{yc_1c_2}\otimes P_{d_1}$ are
replaced with $V_{xa_1\ldots a_N}$, $V_{yc_1\ldots c_N}$,
$P_{xa_1\ldots a_N}\otimes (-1)^t$ and $P_{yc_1\ldots c_N}\otimes
(-1)^q$, respectively.

Same as the case with two agents, Alice need only publish two bits
of classical information for each (N+1)-particle GHZ state
measurement. The controllers are required only to perform two
single-particle measurements along the $x$ direction, $\sigma_x$
for their particles and the receiver can obtain the arbitrary
two-particle state $\vert \Phi\rangle_{xy}$ with two local unitary
operations if she collaborates with the other $N-1$ agents. As all
the quantum source are used to carry the useful information and no
particles are abandoned in this scheme, the intrinsic efficiency
for qubits approaches the maximal value. The security of this QSTS
scheme depends on the process that Alice shares the EPR pairs with
the agents. The ways for sharing a sequence of EPR pairs securely
between two remote men have been discussed in Refs.
\cite{two-step,guoatom,ControlledTele3}. So this QSTS scheme can
be made to be secure.

Quantum state sharing is the extension of quantum secret sharing,
and is used to split an unknown quantum state. For sharing a
classical information, single photons can be used as the quantum
source for setting up the quantum channel
\cite{GuoQSSPLA,dengPLA,Hsu}. For splitting an unknown state, the
quantum source has to be an entangled quantum system. Although big
process has been made for producing entanglement, the efficiency
is still low, in particular for multipartite entanglement
\cite{Mentanglement}. With the present techniques, the EPR pairs
may be one of the optimal entangled quantum sources for quantum
state sharing and quantum teleportation \cite{teleportation}. On
the other hand, the disadvantage of this scheme is that the joint
measurement done by the sender, Alice becomes more difficult with
the increase of the agents. With the development of technology, it
is likely easy for measuring a multipartite entanglement.

In summary, we have presented a way for quantum state sharing of
an arbitrary two-particle state with $2N$ EPR pairs. Any one in
the $N$ agents can recover the original state with two local
unitary operations if he collaborates with the other agents, the
$N-1$ controllers who are required only to perform two
single-particle measurements along the $x$ direction, $\sigma_x$,
without Bell state joint measurements, which makes it more
convenient for the agents in its applications than others.
Certainly, Alice has to perform two multipartite joint
measurements on her particles. Another advantage is that all the
particles can be used to carry the useful information and the
intrinsic efficiency for qubits approaches the maximal value. With
the new notations for GHZ state, Alice need only publish four bits
of classical information for recovering the original state, which
reduces the information exchanged largely.

The author F.G. Deng is very grateful to Dr.  Debbie W. Leung for
her help. This work is supported by the National Natural Science
Foundation of China under Grant Nos. 10447106, 10435020, 10254002
and A0325401.

\begin{widetext}
\begin{center}
\begin{table}[!h]
\label{table1} \caption{The relation between the local unitary
operations and the results $R_{xa_1a_2}$, $R_{yc_1c_2}$, $R_{b_1}$
and $R_{d_1}$. $\Phi_{b_2d_2}$ is the state of the two particles
hold in the hand of Charlie after all the measurements are done by
Alice and Bob; $U_C$ are the local unitary operations with which
Charlie can reconstruct the unknown state $\vert
\Phi\rangle_{xy}$. }
\begin{tabular}{ccccccc|cccccc}\hline
$V_{xa_1a_2}$  & & $V_{yc_1c_2}$& & $P_{xa_1a_2}\otimes P_{b_1}$ &
& $P_{yc_1c_2}\otimes P_{d_1}$ & & & & $\Phi_{b_2d_2}$ & &
$U_C$\\\hline
 0  & & 0 & & $+$ & & $+$ & &  & & $\alpha\vert
00\rangle + \beta\vert 01\rangle + \gamma\vert 10\rangle + \delta
\vert
11\rangle$ & & $U_0\otimes U_0 $ \\
 0  & & 0 & & $+$ & & $-$ & &  & &
$\alpha\vert 00\rangle - \beta\vert 01\rangle + \gamma \vert
10\rangle - \delta \vert 11\rangle$ & & $U_0\otimes U_1 $
\\
0  & & 0 & & $-$ & & $+$ & &  & & $\alpha\vert 00\rangle +
\beta\vert 01\rangle - \gamma\vert 10\rangle - \delta \vert
11\rangle$ & & $U_1\otimes U_0 $
\\
 0  & & 0 & & $-$ & & $-$ & &  & & $\alpha\vert
00\rangle - \beta\vert 01\rangle - \gamma\vert 10\rangle + \delta
\vert 11\rangle$ & & $U_1\otimes U_1 $
\\
 0  & & 1 & & $+$ & & $+$ & &  & & $\alpha\vert
01\rangle + \beta\vert 00\rangle + \gamma  \vert 11\rangle +
\delta\vert 10\rangle$ & & $U_0\otimes U_2 $
\\
 0  & & 1 & & $+$ & & $-$ & &  & & $\alpha\vert
01\rangle - \beta\vert 00\rangle + \gamma\vert 11\rangle - \delta
\vert 10\rangle$ & & $U_0\otimes U_3 $
\\
 0  & & 1 & & $-$ & & $+$ & &  & & $\alpha\vert
01\rangle + \beta\vert 00\rangle - \gamma\vert 11\rangle - \delta
\vert 10\rangle$ & & $U_1\otimes U_2 $
\\
 0  & & 1 & & $-$ & & $-$ & &  & & $\alpha\vert
01\rangle - \beta\vert 00\rangle - \gamma\vert 11\rangle + \delta
\vert 10\rangle$ & & $U_1\otimes U_3 $
\\
 1  & & 0 & & $+$ & & $+$ & &  & & $\alpha\vert
10\rangle + \beta\vert 11\rangle + \gamma\vert 00\rangle + \delta
\vert 01\rangle$ & & $U_2\otimes U_0 $
\\
 1  & & 0 & & $+$ & & $-$ & &  & & $\alpha\vert
10\rangle - \beta\vert 11\rangle + \gamma \vert 00\rangle - \delta
\vert 01\rangle$ & & $U_2\otimes U_1 $
\\
 1  & & 0 & & $-$ & & $+$ & &  & & $\alpha\vert
10\rangle + \beta\vert 11\rangle - \gamma\vert 00\rangle - \delta
\vert 01\rangle$ & & $U_3\otimes U_0$
\\
 1  & & 0 & & $-$ & & $-$ & &  & & $\alpha\vert
10\rangle - \beta\vert 11\rangle - \gamma\vert 00\rangle + \delta
\vert 01\rangle$ & & $U_3\otimes U_1 $
\\
 1  & & 1 & & $+$ & & $+$ & &  & & $\alpha\vert
11\rangle + \beta\vert 10\rangle + \gamma\vert 01\rangle + \delta
\vert 00\rangle$ & & $U_2\otimes U_2 $
\\
 1  & & 1 & & $+$ & & $-$ & &  & & $\alpha\vert
11\rangle - \beta\vert 10\rangle + \gamma\vert 01\rangle - \delta
\vert 00\rangle$ & & $U_2\otimes U_3 $
\\
 1  & & 1 & & $-$ & & $+$ & &  & & $\alpha\vert
11\rangle + \beta\vert 10\rangle - \gamma\vert 01\rangle - \delta
\vert 00\rangle$ & & $U_3\otimes U_2 $
\\
1  & & 1 & & $-$ & & $-$ & &  & & $\alpha\vert 11\rangle -
\beta\vert 10\rangle -\gamma\vert 01\rangle + \delta \vert
00\rangle$ & & $U_3\otimes U_3 $
\\\hline
\end{tabular}
\end{table}
\end{center}
\end{widetext}


\begin{thebibliography}{99}
\bibitem{Blakley} G. R. Blakley, in \textit{Proceedings of the American Federation
of Information Processing 1979 National Computer Conference}
(American Federation of Information Processing, Arlington, VA,
1979), pp.313-317; A. Shamir, Commun. ACM \textbf{22}, 612 (1979).

\bibitem{HBB99} M. Hillery et al., Phys.
Rev. A \textbf{59}, 1829(1999).

\bibitem{KKI} A. Karlsson et al., Phys. Rev. A \textbf{59}, 162 (1999).

\bibitem{longqss} L. Xiao et al., Phys.
Rev. A  \textbf{69}, 052307 (2004); F. G. Deng et al., Chin. Phys.
Lett. \textbf{21}, 2097 (2004); Z. J. Zhang et al., Phys. Rev. A
\textbf{71}, 044301 (2005); Z. J. Zhang et al., Eur. Phys. J. D.
\textbf{33}, 133 (2005).

\bibitem{QSSS} D. Gottesman, Phys. Rev. A \textbf{61}, 042311 (2000);
S. Bandyopadhyay, Phys. Rev. A \textbf{62}, 012308 (2000); C. P.
Yang et al., J. Opt. B \textbf{3}, 407 (2001); V. Karimipour et
al., Phys. Rev. A \textbf{65}, 042320 (2002); S. Bagherinezhad and
V. Karimipour, Phys. Rev. A \textbf{67}, 044302 (2003);  A.
Sen(De) et al., Phys. Rev. A \textbf{68}, 032309 (2003).



\bibitem{GuoQSSPLA} G. P. Guo and G. C. Guo, Phys. Lett. A \textbf{310}, 247
(2003);

\bibitem{dengPLA} F. G. Deng et al., Phys. Lett. A \textbf{337},
 329 (2005);  Phys. Lett. A \textbf{340}, 43 (2005).

\bibitem{Hsu} L.-Y. Hsu  and C. -M. Li, Phys. Rev. A \textbf{71}, 022321
(2005).


\bibitem{cleve} R. Cleve et al.,
Phys. Rev. Lett. \textbf{83}, 648 (1999).

\bibitem{nascimento} A. C. A. Nascimento et al., Phys. Rev. A \textbf{64}, 042311 (2001).

\bibitem{Peng} Y. M. Li et al., Phys. Lett.
A \textbf{324}, 420 (2004).

\bibitem{Lance} A. M. Lance et al., Phys. Rev. A \textbf{71},
033814 (2005).


\bibitem{TZG} W. Tittel et al., Phys. Rev. A \textbf{63}, 042301 (2001).

\bibitem{AMLance} A. M. Lance et al., Phys. Rev. Lett. \textbf{92}, 177903 (2004).



\bibitem{QKDs} N. Gisin et al., Rev. Mod. Phys. \textbf{74}, 145 (2002); G. L. Long and
X. S. Liu, Phys. Rev. A \textbf{65}, 032302 (2002);  W. Y. Hwang,
Phys. Rev. Lett. \textbf{91}, 057901 (2003); F. G. Deng and G. L.
Long,  Phys. Rev. A \textbf{68}, 042315 (2003); Phys. Rev. A
\textbf{70}, 012311 (2004).

\bibitem{QSDC} F. G. Deng and G. L. Long,
Phys. Rev. A \textbf{69}, 052319 (2004); F. L. Yan and X. Q.
Zhang, Euro. Phys. J. B \textbf{41}, 75 (2004); T. Gao et al.,
Nuove Cimento B \textbf{119}, 313 (2004); Z. X. Man et al., Chin.
Phys. Lett. \textbf{22}, 18 (2005); C. Wang et al., Phys. Rev. A,
\textbf{71}, 044305 (2005).


\bibitem{two-step}  F. G. Deng et al., Phys. Rev. A \textbf{68}, 042317 (2003).

\bibitem{ControlledTele} A. Karlsson and M. Bourennane, Phys. Rev. A \textbf{58},
4394 (1998).

\bibitem{ControlledTele2} C. P. Yang et al., Phys. Rev. A \textbf{70},
022329 (2004).

\bibitem{ControlledTele3} F. G. Deng et al,
Phys. Rev. A \textbf{72}, 022338 (2005).

\bibitem{Leung} D. W. Leung, Int. J. Quant. Inform. \textbf{2}, 33
(2004); P. Aliferis and  D. W. Leung, Phy. Rev. A \textbf{70},
062314 (2004).


\bibitem{guoatom} C. P. Yang and G. C. Guo, Phys. Rev. A \textbf{59}, 4217
(1999).

\bibitem{book} M. A. Nielsen and I. L. Chuang, \emph{Quantum computation
and quantum information} (Cambridge University Press, Cambridge,
UK, 2000).

\bibitem{Mentanglement} D. Bouwmeester et al., Phys. Rev. Lett. \textbf{82}, 1345  (1999); J. -W. Pan,
et al., Phys. Rev. Lett. \textbf{86}, 4435 (2001); Z. Zhao et al.,
Nature \textbf{430}, 54 (2004).


\bibitem{teleportation} C. H. Bennett et al., Phys.
Rev. Lett. \textbf{70}, 1895 (1993); D. Bouwmeester et al., Nature
\textbf{390}, 575 (1997); A. Furusawa et al., Science
\textbf{282}, 706 (1998); D. Boschi et al., Phys. Rev. Lett.
\textbf{80}, 1121 (1998); Y. H. Kim, S. P. Kulik, and Y. Shih,
Phys. Rev. Lett. \textbf{86}, 1370 (2001).



\end{thebibliography}
\end{document}